\documentclass[11pt]{article} 
\pdfoutput=1
\usepackage{natbib}
\usepackage{graphicx}
\usepackage{hyperref}
\usepackage{amssymb}
\usepackage{authblk}
\usepackage{aas_macros}
\usepackage{rotating}
\date{\today}

\begin{document} 

\title{ANTARES constrains a blazar origin of two IceCube PeV neutrino events}

\author[1]{S.~Adri\'an-Mart\'inez}
\author[2] {A.~Albert}
\author[3] {M.~Andr\'e} 
\author[5] {G.~Anton} 
\author[1] {M.~Ardid} 
\author[6] {J.-J.~Aubert}
\author[7] {B.~Baret}
\author[8] {J.~Barrios}
\author[9] {S.~Basa}
\author[6] {V.~Bertin}
% ...

\affil[1]{\scriptsize{Institut d'Investigaci\'o per a la Gesti\'o Integrada de les Zones Costaneres (IGIC) - Universitat Polit\`ecnica de Val\`encia. C/ Paranimf 1, 46730 Gandia, Spain.}}
\affil[2]{\scriptsize{GRPHE - Universit\'e de Haute Alsace - Institut Universitaire de Technologie de Colmar, 34 rue du Grillenbreit BP 50568 - 68008 Colmar, France}}
\affil[3]{\scriptsize{Technical University of Catalonia, Laboratory of Applied Bioacoustics, Rambla Exposici\'o, 08800 Vilanova i la Geltr\'u, Barcelona, Spain}}
% ...
\affil[5]{\scriptsize{Friedrich-Alexander-Universit\"at Erlangen-N\"urnberg, Erlangen Centre for Astroparticle Physics, Erwin-Rommel-Str. 1, 91058 Erlangen, Germany}}
\affil[6]{\scriptsize{CPPM, Aix-Marseille Universit\'e, CNRS/IN2P3, Marseille, France}}
\affil[7]{\scriptsize{APC, Universit\'e Paris Diderot, CNRS/IN2P3, CEA/IRFU, Observatoire de Paris, Sorbonne Paris Cit\'e, 75205 Paris, France}}
\affil[8]{\scriptsize{IFIC - Instituto de F\'isica Corpuscular, Edificios Investigaci\'on de Paterna, CSIC - Universitat de Val\`encia, Apdo. de Correos 22085, 46071 Valencia, Spain}}
\affil[9]{\scriptsize{LAM - Laboratoire d'Astrophysique de Marseille, P\^ole de l'\'Etoile Site de Ch\^ateau-Gombert, rue Fr\'ed\'eric Joliot-Curie 38, 13388 Marseille Cedex 13, France}}
% ...

\maketitle

\abstract{The source(s) of the neutrino excess reported by the IceCube Collaboration is unknown. The TANAMI Collaboration recently reported on the multiwavelength emission of six bright, variable blazars which are positionally coincident with two of the most energetic IceCube events. Objects like these are prime candidates to be the source of the highest-energy cosmic rays, and thus of associated neutrino emission. We present an analysis of neutrino emission from the six blazars using observations with the ANTARES neutrino telescope.The standard methods of the ANTARES candidate list search are applied to six years of data to search for an excess of muons --- and hence their neutrino progenitors --- from the directions of the six blazars described by the TANAMI Collaboration, and which are possibly associated with two IceCube events. Monte Carlo simulations of the detector response to both signal and background particle fluxes are used to estimate the sensitivity of this analysis for different possible source neutrino spectra. A maximum-likelihood approach, using the reconstructed energies and arrival directions of through-going muons, is used to identify events with properties consistent with a blazar origin.Both blazars predicted to be the most neutrino-bright in the TANAMI sample (1653$-$329 and 1714$-$336) have a signal flux fitted by the likelihood analysis corresponding to approximately one event. This observation is consistent with the blazar-origin hypothesis of the IceCube event IC14 for a broad range of blazar spectra, although an atmospheric origin cannot be excluded. No ANTARES events are observed from any of the other four blazars, including the three associated with IceCube event IC20. This excludes at a 90\% confidence level the possibility that this event was produced by these blazars unless the neutrino spectrum is flatter than $-2.4$.}

\section{Introduction}

Since the initial report of the observation of two high-energy ($\sim$PeV) neutrino-induced cascades by the IceCube Collaboration \citep{2013PhRvL.111b1103A}, further observations using the high-energy starting-event (HESE) analysis have revealed an excess of events consistent with an isotropic, flavour-uniform flux of astrophysical neutrinos \citep{2013Sci...342E...1I,2014arXiv1410.1749A,2014PhRvL.113j1101A}. The small number of excess events ($37$ total, with an estimated background of $15$), and directional resolution of typically $10^{\circ}$ or worse for cascades, makes it difficult to resolve potential features of this flux, such as a spectral downturn above PeV energies, a steeper spectral index, and/or a contribution from one or more point-like sources of neutrinos. Consequently, many suggestions for the nature and origin(s) of this flux have been put forward. Of particular note is the suggestion of a point-source near the Galactic Centre producing the observed excess in that region \citep{2013PhRvD..88h1302R}, a hypothesis already constrained by the ANTARES Collaboration \citep{2014ApJ...786L...5A}.

The TANAMI Collaboration has recently reported observations of six bright, variable blazars in positional coincidence with the range of possible arrival directions of the two PeV IceCube events IC14 and IC20 \citep{2014A&A...566L...7K}\footnote{The paper was released before the third PeV event, IC35 (`Big Bird'), was made public.  A search for possible blazar associations with this event is in preparation by the TANAMI Collaboration.}. Using a simple calculation based on the observed 1~keV to 10~GeV photon flux, the authors estimate that $1.9 \pm 0.4$ electron neutrino events at PeV energies would be expected in $662$ days of IceCube data. This estimate compares well with the two observed events IC14 and IC20. Even taking this only as an order-of-magnitude indication of the expected event rate, a higher-resolution follow-up study of these objects is of great interest. Here, we present such an analysis using six years of data from the ANTARES neutrino telescope.

\section{Target blazars and possible neutrino fluxes}
\label{SecBlazarNu}

\begin{sidewaystable}
{\small
\centering
\begin{tabular}{llccccccc}
\hline \hline								
Source & Cat. Name & R.A. & Dec. & Class & $z$ & $F_{\gamma}$ & $N_{\nu_e}$ & IC \\
	&	& [$^{\circ}$] & [$^{\circ}$] & && [GeV cm$^{-2}$ s$^{-1}$] & & \\
\hline
0235$-$618 &  PKS\,0235$-$618 & 39.2218 & $-$61.6043 & Q & 0.47$^{\textrm{a}}$ &$\left( 6.2^{+3.1}_{-3.1} \right) \times 10^{-8}$ & $0.19^{+0.04}_{-0.04}$ & 20, 7\\
0302$-$623 & PKS\,0302$-$623 & 45.9610 & $-$62.1904 & Q & 1.35$^{\textrm{a}}$ &$\left( 2.1^{+0.4}_{-0.4} \right) \times 10^{-8}$ & $0.06^{+0.01}_{-0.01}$ & 20\\
0308$-$611 & PKS\,0308$-$611 & 47.4838 & $-$60.9775 & Q & 1.48$^{\textrm{a}}$ &$\left( 4.7^{+1.8}_{-1.8} \right) \times 10^{-8}$ & $0.14^{+0.05}_{-0.05}$ & 20\\
\hline
1653$-$329 & Swift\, J1656.3$-$3302 & 254.0699 & $-$33.0369 & Q & 2.40$^\textrm{b}$ &$\left( 2.8^{+0.3}_{-0.3} \right) \times 10^{-7}$ & $0.86^{+0.10}_{-0.10}$ & 14, 2, 25\\
1714$-$336 & TXS\,1714$-$336  & 259.4001 & $-$33.7024 & B/Q & ? &$\left( 1.5^{+0.3}_{-0.4} \right) \times 10^{-7}$ & $0.46^{+0.10}_{-0.12}$ & 14,2,25\\
1759$-$396 & MRC\,1759$-$396  & 270.6778 & $-$39.6689 & Q & 1.32$^\textrm{c}$ &$\left( 7.5^{+1.9}_{-1.9} \right) \times 10^{-8}$ & $0.23^{+0.50}_{-0.40}$ & 14, 2, 15, 25\\
\hline
\end{tabular}}
\caption{Basic data on the six blazars studied in this analysis. Columns: (1) IAU B1950 name; (2) Common catalog name; (3,4) J\,2000 coordinates; (5) Classification: Q -- Flat Spectrum Radio Quasar, B -- BL\,Lac object; (6) Redshift: $^{\textrm{a}}$ \cite{2008ApJS..175...97H}, $^{\textrm{b}}$ \cite{2003yCat.2246....0C}, $^{\textrm{c}}$ \cite{2009A&A...495..691M}; (7) Total high-energy photon flux from \citet{2014A&A...566L...7K}; (8) Estimated number $N_{\nu_e}$ of $\nu_e$ events in the IceCube $662$-day analysis \citep{2013Sci...342E...1I}; (9) IC gives the IceCube event IDs from \citet{2014PhRvL.113j1101A} with which the blazars are positionally consistent within the angular error range from \citet{2013Sci...342E...1I}.}  \label{TabAGN}
\end{sidewaystable}

The six blazars associated with the IC14 and IC20 fields by \citet{2014A&A...566L...7K} are listed in Table \ref{TabAGN}. All exhibit prominent high-energy photon emission, and all but one are classified as flat-spectrum radio quasars (FSRQs) \citep{2006A&A...455..773V}.
 The predictions of the expected number of detected electron neutrino events were made by assuming a neutrino energy $E_{\nu}=1$~PeV and a flavour-uniform flux, with total energy flux equal to that in high-energy photons.
Active galactic nuclei (AGN) of all classes have long been proposed as sites of hadronic interaction, and are potential sources of the highest-energy cosmic rays and, hence, neutrinos \citep{1975Ap&SS..32..461B,1984ARA&A..22..425H,1996SSRv...75..341S,2014MNRAS.443..474P}. Predictions for the neutrino flux depend on the nature of the AGN considered, the cosmic-ray composition and flux, and the assumed densities of target hadronic matter and magnetic and photon fields \citep{1994APh.....2..375S,1995APh.....3..295M,1999PhRvD..59b3002W,2001PhRvL..87v1102A,PhysRevD.74.034018,2014PhRvD..89l3005B, 2014JHEAp...3...29D}.

The emphasis on the two PeV events \citep[IC14 and IC20; see ][for a full list]{2014PhRvL.113j1101A} comes from the fact that these two highest-energy events have only a negligible probability for an atmospheric origin. While IC14 and IC20 are assumed to be $\nu_e$ charged-current (CC) events, and the most common production mechanism (photo-pion production) produces a flux which is almost uniform in neutrino flavour at Earth, a flavour-dependent flux is predicted by different initial neutrino production mechanisms \citep{2014PhRvD..90l3006K,2015PhRvD..91b7301A}, and/or by invoking new physics during propagation \citep[and references therein]{2003PhRvD..68i3005B}.

 The IceCube observations allow for the possibility of a sub-PeV flux of neutrinos from the sample blazars, in that four other events are positionally associated with the blazar sample (see Table \ref{TabAGN}). This is also consistent with the prediction of two $\nu_e$ charged-current (CC) events, since the low flavour-dependence of the IceCube HESE effective area at the highest energies means an equal number of $\nu_{\mu}$ and $\nu_{\tau}$ events would be expected from a flavour-uniform flux, but with a lower deposited energy. IceCube data are currently compatible with a flavour-uniform flux above $35$~TeV \citep{2015arXiv150203376I}, but a significant excess or deficit of track-like (mostly $\nu_{\mu}$ CC) events in the cosmic diffuse flux cannot be excluded. Thus while these additional four events do not represent a significant excess above a diffuse background, the possibility that they may originate from the blazars in question should also be tested.

\section{ANTARES candidate list search and expected sensitivity}
\label{SecAntares}

ANTARES is an underwater neutrino telescope located in the Mediterranean Sea off the coast of Toulon, at $42^{\circ}48'$~N, $6^{\circ}10'$~E \citep{2011NIMPA.656...11A}. Consisting of an array of photomultiplier tubes, it is designed to record the induced Cherenkov light from the passage of energetic charged particles to infer the interactions of neutrinos.

The ANTARES candidate list search (CLS) methodology is described in \citet{2012ApJ...760...53A}, with the latest results using six years of data ($1338$ days effective livetime) presented in \citet{2014ApJ...786L...5A}. The search uses only up-going muons (i.e., those originating from below the horizon), with cuts placed on the fit quality of the muon track reconstruction and the estimated angular error. The long range of relativistic muons in seawater and the Earth's crust extends the effective detection volume to well beyond the physical size of the detector, in contrast with a HESE-like analysis. The six-year sample consists of $5516$ events, with an estimated atmospheric muon contamination of $10$\%, and an estimated median angular resolution of $0.38^{\circ}$. A maximum-likelihood method is then used to estimate the relative contributions of signal and background fluxes, based on both the reconstructed event arrival directions and the fitted number of photon hits (a robust proxy for energy). We note that this method results in a non-integer number of signal events $N_{\rm sig}$ being estimated, since the signal and background fluxes maximising the likelihood of a given observation can take any normalisation. We also note that it is optimised assuming an $E_{\nu}^{-2}$ source spectrum, and it is sensitive almost exclusively to muon neutrinos. The ability of the ANTARES CLS to constrain the origin of the IceCube events is therefore dependent on the flavour ratio, which may vary according to the neutrino-production scenarios discussed in Sect.\ \ref{SecBlazarNu}. Hereafter, sensitivities and limits will be shown for a uniform flavour ratio, from which results for non-uniform flavour fluxes can readily be derived.

\begin{figure*}
\centering
\includegraphics[width=0.5 \textwidth]{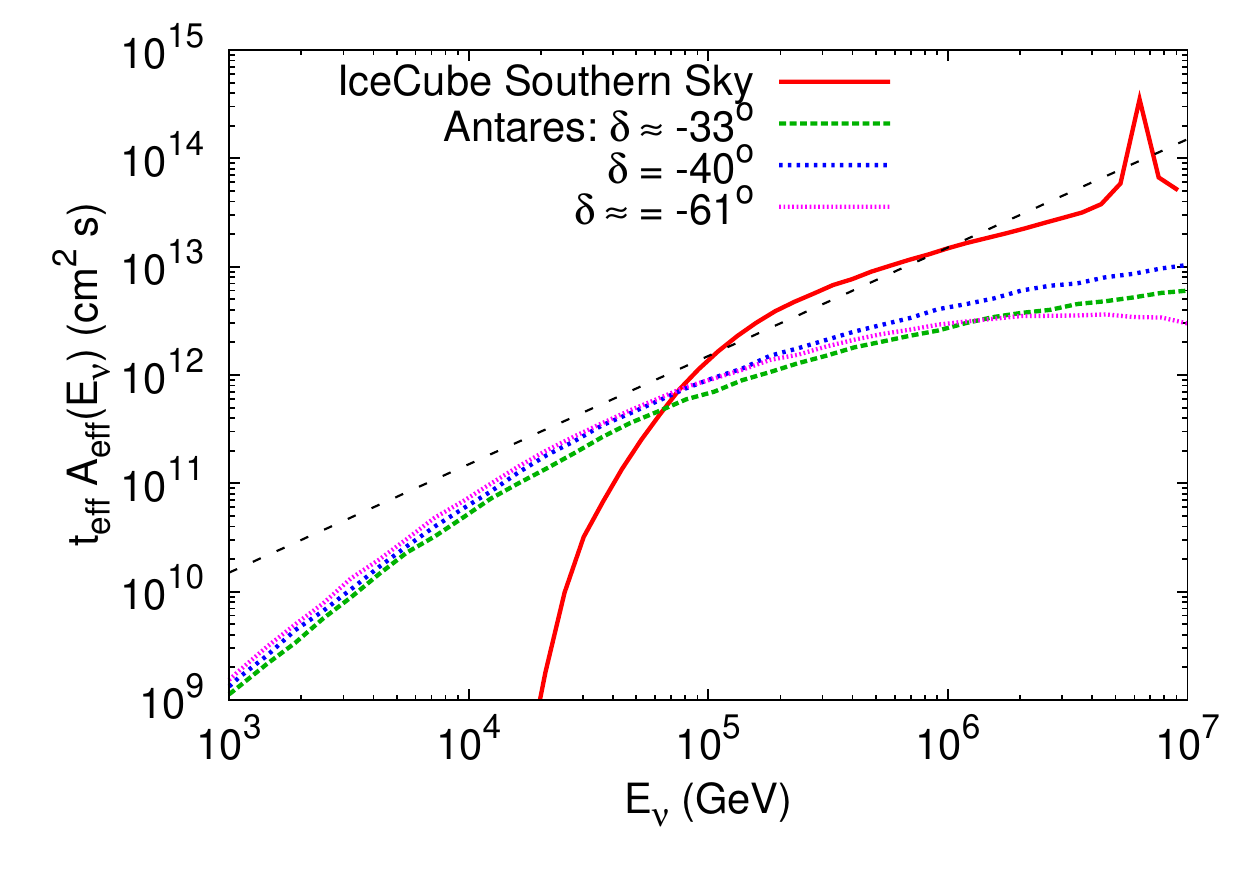}  \includegraphics[width=0.49 \textwidth]{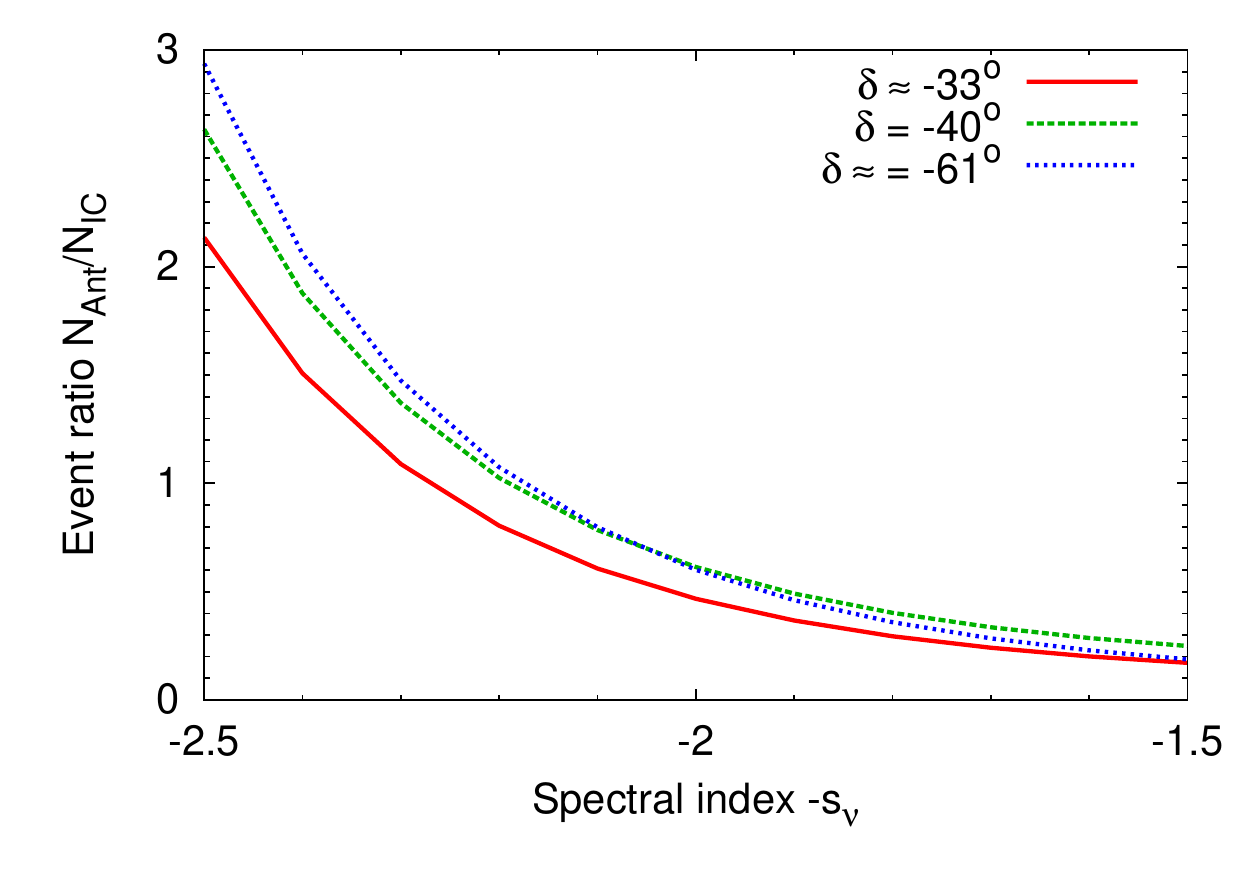} 
\caption{
(left)
Relative exposures of the ANTARES CLS \citep{2012ApJ...760...53A} to a flavour-uniform neutrino flux from the characteristic declinations of the six candidate blazars, and the Southern-Sky-average of the IceCube HESE analysis \citep{2014PhRvL.113j1101A} (exposures from \citet{2013Sci...342E...1I}). The black dashed line, included for reference, is proportional to $E_{\nu}$.
(right) Expected number of ANTARES events per detected IceCube event for power-law spectra (Eq.\ \ref{EqSpectrum}) as a function of the neutrino spectral index $-s_{\nu}$, calculated using the relative exposures.
}
\label{FigRelExposure}
\end{figure*}

The ability of the ANTARES CLS to probe the PeV-neutrino blazar-origin hypotheses of \citet{2014A&A...566L...7K} can be seen from Fig.\ \ref{FigRelExposure}, which compares the time-integrated, flavour-averaged exposures of the ANTARES CLS (\citet{2014ApJ...786L...5A}; $1338$ days, using one third of the effective area to muon neutrinos) at the characteristic declinations of the six blazars considered here, to that of the IceCube HESE analysis, averaged over the southern hemisphere (\citet{2013Sci...342E...1I}; now updated to 998 days by \citet{2014PhRvL.113j1101A}, averaged over all three neutrino flavours). It can be seen that below approximately $100$~TeV, ANTARES has a greater sensitivity to a neutrino flux from the six blazars at the given southern declinations than the recent IceCube HESE analysis.

The predictions for the number of IceCube-detected PeV  neutrino events by \citet{2014A&A...566L...7K} were based on equating the neutrino flux at $1$~PeV to the integrated photon flux between $1$~keV and $10$~GeV. While the expected neutrino-flux shape is highly model-dependent (as was discussed in Sect.\ \ref{SecBlazarNu}), the prediction that the total neutrino energy flux $F_{\nu}$ (GeV~cm$^{-2}$~s$^{-1}$) is approximately equal to the total high-energy photon flux $F_{\gamma}$ is relatively robust, at least when attributing this emission to a $100$\% hadronic origin. The black-dashed line in Fig.\ \ref{FigRelExposure} is proportional to neutrino energy $E_{\nu}$ and normalised to the IceCube exposure at $1$~PeV, i.e., it is a line of equal sensitivity to a neutrino flux $F_{\nu}$. For constant $F_{\nu}$, it is clear that the IceCube HESE analysis is most sensitive to a flux at a few hundred TeV, while the ANTARES CLS is most sensitive near $30$~TeV.

\begin{figure}
\centering
\includegraphics[width=0.49 \textwidth]{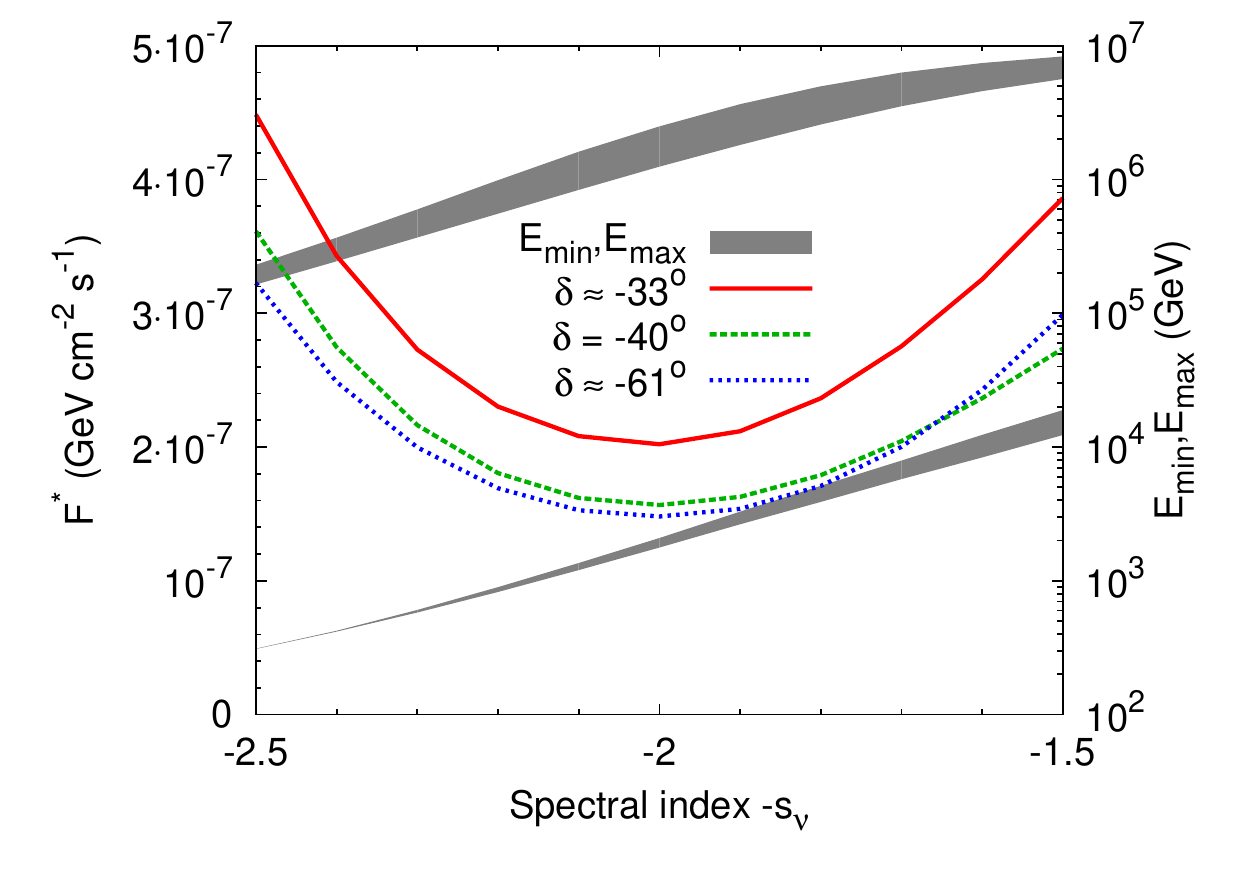}
\caption{Neutrino flux $F_{\nu}^*$ required to produce one neutrino event in ANTARES as a function of spectral index $s_{\nu}$ (Eq.\ \ref{Eqnfstar}). The corresponding energy ranges of integration $E_{\rm min}$ and $E_{\rm max}$ (Eq.\ \ref{EqEminmax}) are shown as lower and upper shaded regions respectively: the shading covers the variation due to declination.}
\label{FigFluxes}
\end{figure}

The range of potential neutrino spectra, $\Phi_{\nu}(E_{\nu})$ ($dN_{\nu}/dE_{\nu}$), are characterised by generic power-law spectra with spectral index $-s_{\nu}$:
\begin{eqnarray}
\Phi_{\nu}(E_{\nu}) & = & \Phi_0 \left( \frac{E_{\nu}}{1~{\rm GeV}} \right)^{-s_{\nu}}~~~{\rm [GeV^{-1}~cm^{-2}~s^{-1}]}. \label{EqSpectrum}
\end{eqnarray}
The relative numbers of events expected to be observed by ANTARES compared to IceCube for such spectra are shown in Fig.\ \ref{FigRelExposure} (right). The required energy in such fluxes to produce a single detectable event in ANTARES is calculated in Appendix \ref{sec:energy_flux}, and plotted in Fig.\ \ref{FigFluxes} (online only). In the range $-2.5 < -s_{\nu} < -1.5$ , it is comparable with the total blazar photon flux calculated by \citet{2014A&A...566L...7K} (see Table \ref{TabAGN}).

Having established a wide range of plausible neutrino flux scenarios, and the sensitivity of the ANTARES CLS to neutrino fluxes over a broad range of energies, we therefore perform the standard ANTARES CLS for an excess of neutrino emission from the six candidate blazars.

\section{Results and discussion}
\label{SecResults}

\begin{table}
{\small
\label{TabResults}
\centering
\begin{tabular}{lcccc}
\hline \hline
Source      & $N_{\rm sig}$ & $p$  & Limit & $N_{\nu,IC}=1, 2, 3, 4$ \\
\hline
0235$-$618  &  0 & 1  & 1.3 & -2.4 ~ -2.1 ~ -2.0 ~ -1.9 \\
0302$-$623  &  0 & 1  & 1.3 & -2.4 ~ -2.1 ~ -2.0 ~ -1.9 \\
0308$-$611  &  0 & 1  & 1.3 & -2.4 ~ -2.1 ~ -2.0 ~ -1.9 \\
\hline
1653$-$329  &  1.1 & 0.10  & 2.9 & $<$-2.5 ~ -2.5 ~ -2.3 ~ -2.2 \\
1714$-$336  &  0.9 & 0.04  & 3.5 & $<$-2.5 ~ -2.5 ~ -2.3 ~ -2.2 \\
1759$-$396  &  0   & 1     & 1.4 & -2.4 ~ -2.1 ~ -2.0 ~ -1.8\\
\hline
\end{tabular}}
\caption{ANTARES point-source analysis results. Columns: (1) IAU B\,1950 name; (2) Number of fitted signal events; (3) pre-trial $p$-value;  (4) $90$\% upper limit ($10^{-8}$~GeV$^{-1}$~cm$^{-2}$~s$^{-1}$) on $\Phi_0$ for $-s_{\nu}=-2.0$, (5): minimum spectral indices $-s_{\nu}$ consistent at $90$\% C.L.\ with $N_{\nu,IC}=1$\ldots$4$ associated IceCube events. Limits assume a flavour-uniform flux.}
\end{table}

The results of the ANTARES analysis of the six blazars are given in Table \ref{TabResults}. For four of the six targets, no source-like neutrinos were identified ($N_{\rm sig}=0$), allowing relatively strong upper limits to be placed on an $E_{\nu}^{-2}$ flux. Blazars 1653$-$329 and 1714$-$336 were each fitted as having approximately one nearby signal-like event, with $N_{\rm sig}$ of $1.1$ and $0.9$ respectively\footnote{The maximum-likelihood procedure estimates $N_{\rm sig}$ as a continuous variable, as discussed in Sect.\ \ref{SecAntares}.}. This observation is well within the expected background fluctuations, however, with pre-trial $p$-values (probability of the likelihood procedure fitting a stronger signal flux to background-only data) of $0.10$ and $0.04$, respectively\footnote{The correct penalty factor for multiple trials is $61$, including the six blazars considered here, and $55$ trials from other analyses using the CLS \citep{2014ApJ...786L...5A,2014arXiv1407.8525A}}.  Nonetheless, it must be noted that these two blazars have the highest predicted neutrino fluxes, and that from Fig.\ \ref{FigRelExposure} (right), neutrino fluxes with spectral indices between $-2.5$ and $-2.3$ producing one IceCube event would be expected to produce between one and two ANTARES events. Therefore, when the calculation of \citet{2014A&A...566L...7K} is extended to include power-law neutrino spectra, the result of the analysis is consistent with the sample blazars being neutrino sources with fluxes in proportion to their observed high-energy photon flux ($F_{\gamma}$ in Table \ref{TabAGN}), even if the result is also consistent with background.

\begin{figure*}
\centering
\includegraphics[width=0.49 \textwidth]{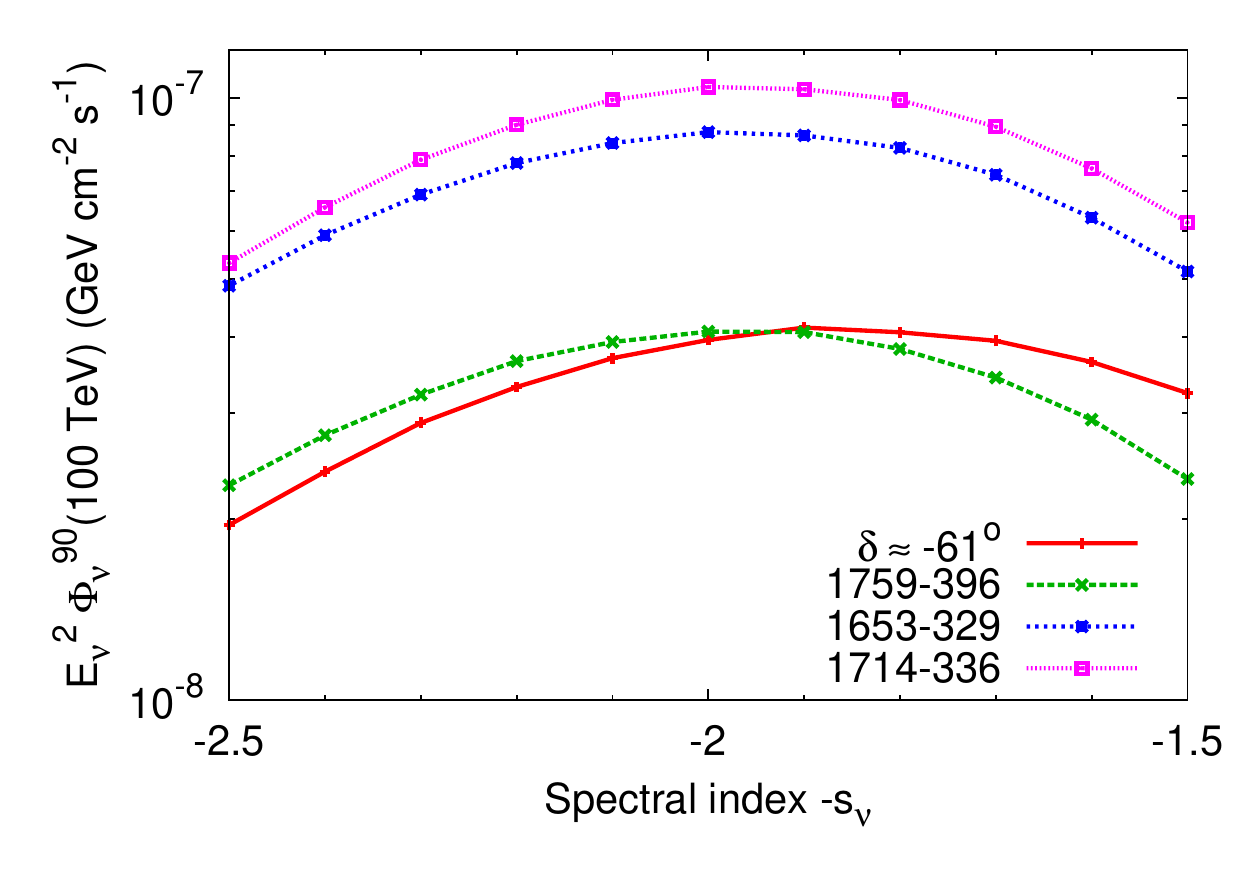}
\includegraphics[width=0.49 \textwidth]{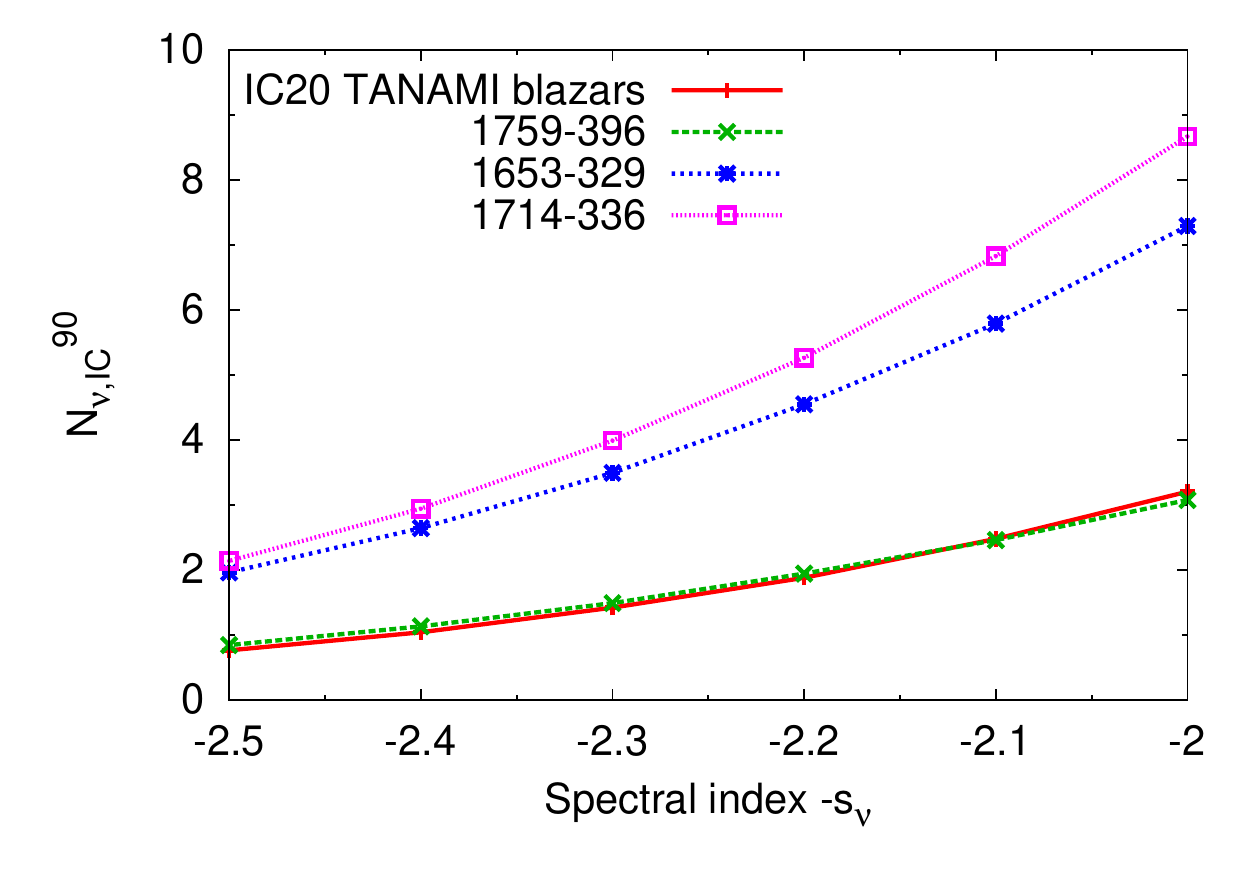}
 \caption{(left)ANTARES 90\% confidence limits on a flavour-uniform neutrino flux ($\Phi_{\nu} \equiv \Phi_{\nu_e} + \Phi_{\nu_{\mu}} + \Phi_{\nu_{\tau}} = 3 \Phi_{\nu_{\mu}}$) from the six blazars as a function of spectral index $s_{\nu}$ (Eq.\ \ref{EqSpectrum}), and (right) corresponding limits on the expected number of IceCube events of blazar origin, using the exposures shown in Fig.\ \ref{FigRelExposure} and the limiting fluxes. Since the limits from 0235$-$618, 0302$-$623, and 0308$-$611 are almost identical, and since no events were observed, the limits also apply to the summed flux from all three of these blazars, and hence only one line is shown, and is labelled `IC20 TANAMI blazars'.}
 \label{FigFluxLimits}
\end{figure*}

Limits at a $90$\% confidence level (C.L.), $\Phi_{\nu}^{90}$, on the spectra from Eq.\ \ref{EqSpectrum} are generated from the ANTARES observations as a function of $s_{\nu}$ over the approximate predicted range (between $1.5$ and $2.5$), using the method of \citet{1937RSPTA.236..333N}. All are upper limits and are given in Fig.~\ref{FigFluxLimits} (left) (online only). The confidence level is given at $100$~TeV, because it is both the approximate energy at which the ANTARES and IceCube analyses have equal exposures and where the flux limit is least sensitive to $s_{\nu}$.

The flux limits correspond to a maximum expected number $N_{\nu,IC}^{90}$ of events observed by IceCube; where this number is less than the observed number of events, a blazar origin can be excluded at 90\% C.L. This is shown in Fig.~\ref{FigFluxLimits} (right) (online only). Any given number of IceCube events is therefore only consistent with a blazar origin for neutrino spectral indices flatter than certain value; minimum values of $-s_{\nu}$ are given for $1$--$4$ events in Table \ref{TabResults} and should be compared to the possible associations in Table \ref{TabAGN}. For the IC14 field for instance, the possibility that blazar 1759$-$396 could be responsible for three or more associated IceCube events is excluded at $90$\% confidence for neutrino spectra steeper than $-2.1$. For spectra steeper than $-2.4$, we can exclude that 1759$-$396 is responsible for any IceCube events. The limits for 1653$-$329 and 1714$-$336 are weaker because of a possible physical association with the two signal-like ANTARES events. Regardless of the association, we can rule out the possibility that the cluster IC14, IC2, and IC25 arose from a single considered blazar with a spectrum steeper than $-2.4$.
For the IC20 grouping, the non-observation of any event from the three candidate blazars means that the $\delta \approx -61^{\circ}$ limit applies both to the individual blazars, and the group as a whole. Therefore, ANTARES observations can rule out a neutrino spectrum steeper than $-2.2$ as being responsible for both IC20 and IC7, and a neutrino spectrum steeper than $-2.4$ being responsible for only one of them. That is, if IC20 does indeed originate from the three associated TANAMI blazars, the neutrino spectral index must be flatter than $-2.4$.

\section{Conclusion}

We have tested the hypothesis of \cite{2014A&A...566L...7K} that the first two PeV neutrino events observed by IceCube, IC14 and IC20, are of blazar origin, by performing a candidate list search (CLS) for an excess muon neutrino flux from the six suggested blazars using six years of ANTARES data.
We are not able to either confirm or rule out a blazar origin of these events, although constraints have been placed on the range of source spectra which could have produced them, particularly in the case of IC20. These constraints assume that muon neutrinos constitute one third of the neutrino flux, and strengthen or weaken proportionally with this fraction. While approximately two ANTARES events were fitted as being more signal-like than background-like by the maximum-likelihood analysis, such a result is completely within the expected background fluctuations, with pre-trial $p$-values of $10$\% and $4$\% for the blazars in question (1653$-$329 and 1714$-$336). It is interesting to note that these two blazars were predicted by \citet{2014A&A...566L...7K} to have the strongest neutrino flux, and that such a result is within expectations for the ANTARES event rate for an $E_{\nu}^{-2}$ to $E_{\nu}^{-2.3}$ neutrino spectrum given that IceCube observes two such events, and $E_{\nu}^{-2.3}$ to $E_{\nu}^{-2.5}$ for a single event of blazar origin. Given these considerations, the TANAMI candidate blazars should be included in all future analyses.

\section*{acknowledgement} {\small
The authors would like to thank A.~Kappes for helpful discussions regarding the IceCube analysis.
The ANTARES authors acknowledge the financial support of the funding agencies:
Centre National de la Recherche Scientifique (CNRS), Commissariat \`a
l'\'ene\-gie atomique et aux \'energies alternatives (CEA), 
Commission Europ\'eenne (FEDER fund and Marie Curie Program),
R\'egion Alsace (contrat CPER), R\'egion
Provence-Alpes-C\^ote d'Azur, D\'e\-par\-tement du Var and Ville de La
Seyne-sur-Mer, France; Bundesministerium f\"ur Bildung und Forschung
(BMBF), Germany; Istituto Nazionale di Fisica Nucleare (INFN), Italy;
Stichting voor Fundamenteel Onderzoek der Materie (FOM), Nederlandse
organisatie voor Wetenschappelijk Onderzoek (NWO), the Netherlands;
Council of the President of the Russian Federation for young
scientists and leading scientific schools supporting grants, Russia;
National Authority for Scientific Research (ANCS), Romania; Ministerio
de Ciencia e Innovaci\'on (MICINN), Prometeo of Generalitat Valenciana
and MultiDark, Spain; Agence de l'Oriental and CNRST, Morocco. We also
acknowledge the technical support of Ifremer, AIM and Foselev Marine
for the sea operation and the CC-IN2P3 for the computing facilities.
The TANAMI authors acknowledge support and partial funding by the Deutsche
  Forschungsgemeinschaft grant WI 1860-10/1 (TANAMI) and GRK 1147,
  Deutsches Zentrum f\"ur Luft- und Raumfahrt grant
  50\,OR\,1311/50\,OR\,1303/50\,OR\,1401, the Spanish MINECO project
  AYA2012-38491-C02-01, the Generalitat
  Valenciana project PROMETEOII/2014/057, the COST MP0905
  action ``Black Holes in a Violent Universe''
and the Helmholtz Alliance for Astroparticle Physics (HAP).
}
 
\begin{appendix}

\section{Calculation of neutrino energy flux}
\label{sec:energy_flux}

For each spectral index $-s_{\nu}$ and source declination $\delta$, the required neutrino flux $\Phi_{\nu}^*(E_{\nu},\delta)$ expected to produce a single ANTARES event can be found from the expression
\begin{eqnarray}
\int_{0}^{\infty} t_{\rm eff} \, A_{\rm eff}(E_{\nu}) \, \Phi_{\nu}^*(E_{\nu},\delta) \, d E_{\nu} ~ = ~ 1,
\end{eqnarray}
where $A_{\rm eff}(E_{\nu},\delta)$ and $t_{\rm eff}$ are respectively the ANTARES effective area and the observation time. While the total energy in such a flux is infinite, the energy over the sensitive range of ANTARES can be calculated by defining characteristic energies $E_{\rm min}(\delta, s_{\nu})$ and $E_{\rm max}(\delta, s_{\nu})$ such that:
\begin{eqnarray}
\int_{E_{\rm min}}^{E_{\rm max}} t_{\rm eff} \, A_{\rm eff}(E_{\nu},\delta) \, \Phi_{\nu}^*(E_{\nu},\delta) \, d E_{\nu} ~=~ 0.9, \label{EqEminmax}
\end{eqnarray}
with $0.05$ below $E_{\rm min}$ and $0.05$ above $E_{\rm max}$. The total neutrino energy flux $F_{\nu}^*(\delta, s_{\nu})$ in the range $E_{\rm min} \le E_{\nu} \le E_{\rm max}$ required to produce one event can then be calculated from $\Phi_{\nu}^*(E_{\nu},\delta)$ as:
\begin{eqnarray}
F_{\nu}^*(\delta, s_{\nu}) ~=~ \frac{1}{0.9} \int_{E_{\rm min}}^{E_{\rm max}} \Phi_{\nu}^*(E_{\nu},\delta) \, E_{\nu} \, d E_{\nu}~~{\rm [GeV~cm^{-2}~s^{-1}]}. \label{Eqnfstar}
\end{eqnarray}
In Fig.\ \ref{FigFluxes} $F_{\nu}^*(\delta, s_{\nu})$ is plotted along with $E_{\rm min}$ and $E_{\rm max}$.

\end{appendix}

\end{document}